\documentclass[journal,letterpaper]{IEEEtran}
\IEEEoverridecommandlockouts
\usepackage{amsmath}
\usepackage{amsfonts}
\usepackage{graphicx}
\usepackage{enumerate}
\usepackage{subfigure}
\usepackage{mathtools}
\usepackage{color}
\ifCLASSINFOpdf
\else
\fi

\begin{document}
\title{Video Processing on the Edge for Multimedia IoT Systems}
\author{\IEEEauthorblockN{Yang Cao, Zeyu Xu, Peng Qin, and Tao Jiang}
\thanks{ \emph{
Yang Cao, Zeyu Xu and Tao Jiang (corresponding author) are with School of Electronic Information and Communications, Huazhong University of Science and Technology, Wuhan, 430074, China (e-mail: \{ycao, xzyxzyxzy\}@hust.edu.cn, Tao.Jiang@ieee.org).}}
\thanks{ \emph{
Peng Qin is with CAEIT, Beijing, 100041, China (e-mail:125529995@qq.com)}
}}


\maketitle
\IEEEpeerreviewmaketitle

\begin{abstract}
In this article, we first survey the current situation of video processing on the edge for multimedia Internet-of-Things (M-IoT) systems in three typical scenarios, i.e., smart cities, satellite networks, and Internet-of-Vehicles. By summarizing a general model of the edge video processing, the importance of developing an edge computing platform is highlighted. Then, we give a method of implementing cooperative video processing on an edge computing platform based on light-weighted virtualization technologies. Performance evaluation is conducted and some insightful observations can be obtained. Moreover, we summarize challenges and opportunities of realizing effective edge video processing for M-IoT systems.
\end{abstract}


\IEEEpeerreviewmaketitle

\section{Introduction}

During the last decade, Internet-of-Things (IoT) has been evolving to a paradigm that enables the interconnection of physical objects (e.g., sensors, cameras, vehicles, and robots) and human in smart cities/homes/factories, and so forth~\cite{CAO2016}. As an emerging type of IoT, multimedia IoT (M-IoT) systems integrate image processing, computer vision and communication networking capabilities, and have the potential to be used in surveillance (e.g., fire/crime detection), remote sensing (e.g., high-speed object tracking) and driving assistance.

Generally speaking, there are two conventional video processing methods, the first one is to preprocess source video chunks\footnote{Video chunk is the minimum video processing unit that contains one or multiple video frames.} at the camera node. Video preprocessing, such as extracting features from video frames, would reduce the amount of data to be delivered to the remote IoT server when a transition from the pixel domain to the feature domain is performed. The original source video can be delivered to the IoT server later if needed. The second one is to directly transmit video chunks to the remote IoT server for processing. However, the measurements in \cite{RBB2016} demonstrate that the above two methods would lead to significant delays. The reason lies in the fact that the limited computational resources of video source nodes might incur computational delay when preprocessing video chunks locally. Meanwhile, the delivery of original video chunks to the remote IoT server can result in network congestions and delays due to the long-distance or multi-hop transmissions. Thus, it is significant to develop other methods that accommodate the requirement of delay-sensitive video processing tasks for M-IoT systems.

Edge computing and its related concepts (e.g., fog computing and cloudlet) can serve as enablers of distributed computing for the preprocessing of video chunks. By leveraging the edge computing technique, redundant computation, storage and communication capabilities of multiple network-edge nodes (e.g., smart phone, smart car, access point, femto-base station, and robot) in the proximal can be utilized through local high-speed wireless/optical networks~\cite{CLJ2016}. Thus, the service delay can be reduced when the video processing task is handled at nearby edge nodes. Moreover, delay-sensitive video processing tasks can be divided into sub-tasks and preprocessed by multiple edge nodes in parallel, to further accelerate the computation~\cite{LCJ2017}.
%
%
%

In edge computing-enabled networks, the handling of video processing tasks for M-IoT systems is different from that in conventional networks~\cite{CHM2016}. For example, the feature extraction from source video chunks is executed at edge nodes instead of at the remote IoT server. Thus, the compression of video chunks, which may degrade the object detection accuracy, is no longer a severe problem since the capacity of local networks is sufficient and lossy compression is not needed. Instead, new problem arises that how to properly assign video sub-tasks to multiple edge nodes. Some recent studies consider video feature extraction performed by multiple edge nodes to improve the system performance. The authors in \cite{ZFL2017} proposed a scheme to assign feature extraction tasks to mobile devices through their web browsers, while taking into account different capabilities of different edge nodes. The authors in \cite{EDF2017} studied the problem of minimizing the completion time of multiple feature extraction tasks that share the communication and computational resources of multiple edge nodes for task offloading. However, all above studies focused on specific video preprocessing task, such as feature extraction, and lacked practical implementation of various video processing functions at diverse edge nodes in a cooperative manner.
%
%
As motivated, the main focus of this article is to develop a flexible edge computing platform that implements edge video processing functions especially for the M-IoT systems.

The main contributions of this article can be summarized as follows:
\begin{itemize}
  \item We thoroughly survey features and requirements of edge video processing for M-IoT systems in different typical scenarios.
  \item We propose a light-weighted virtualization based edge computing platform, which is environment-aware and fully supports cooperative processing across different network edge devices. Performance evaluation reveals that the cooperative video processing is beneficial.
  \item We provide insights about challenges and opportunities of edge video processing for M-IoT systems.
\end{itemize}

The remainder of this article is organized as follows.
First we overview edge video processing features and requirements in M-IoT systems under different scenarios. Next, we introduce the proposed edge computing platform based on light-weighted virtualization technology and conduct performance evaluations. Then, we summarize challenges and opportunities of M-IoT systems. Finally, we conclude our work.

\section{Edge Video Processing for M-IoT systems}

\begin{table*}[!t]
\renewcommand{\arraystretch}{1.3}
\caption{The comparison among M-IoT systems in typical scenarios} \label{tb_1} \centering
\begin{tabular*}{17cm}{c c c c c}
\hline
 Scenario  &  Typical Task     &       Frame Width (in pixels)     &   Main Concern   & Edge Processing Function@Edge Node  \\
\hline\hline
  Smart cities & Object/event detection & $10^3$ level & Low false-positive/negative & Feature extraction@Smart phone   \\
  \hline
  Satellite networks & Target tracking  & $10^4$ level & High PSNR/SSIM & ROI slicing@Satellite processor   \\
  \hline
  Internet-of-Vehicles & Driving assistance & $10^3$ level & Low delay & View transformation@Vehicle OBU \\
  \hline
\end{tabular*}
\end{table*}

In this section, we survey current situation of edge video processing for M-IoT systems in three typical scenarios, namely, smart cities, satellite networks, and Internet-of-Vehicles, which have attracted much attention from both industry and academia. In TABLE~\ref{tb_1}, we briefly summarize the comparison among M-IoT systems in typical scenarios. Besides, general model of the edge video processing is provided.

\subsection{Surveillance in Smart Cities}
\begin{figure}[!t]
  \centering
  \includegraphics[width=0.95\linewidth]{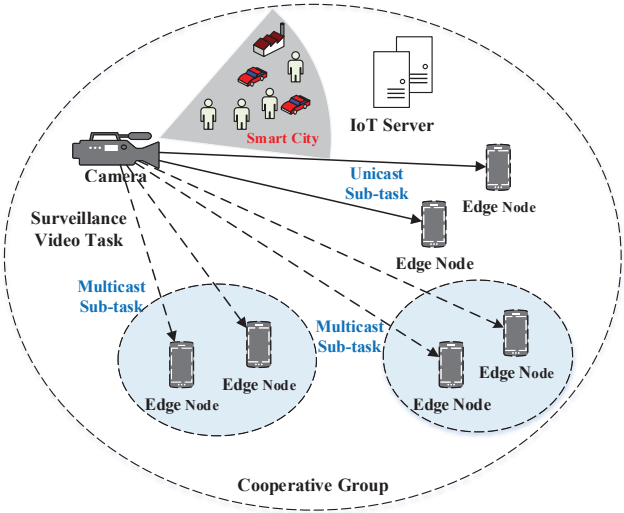}\\
  \caption{Edge computing-assisted surveillance in smart cities.}\label{Fig1}
\end{figure}
A smart city is an ultra-modern urban area that utilizes different IoT technologies to address the diverse needs of businesses, institutions, and citizens. The implementation and deployment of IoT services (e.g., smart transportation, smart offices, smart hospitals, and so on) will improve the quality of life of citizens and the economy. Surveillance and object/event detection through networked static or mobile cameras is essential to the realization of smart city services~\cite{CHM2016}. For example, automatic fire detection can alert the fire department at the earliest time to avoid severe consequence. Crime detection can timely inform the police station to prevent tragedies and property loss. Generally speaking, the process of object/event detection can be divided into two steps, the first step is to extract features from captured video frames, and the second step is to perform specific computation towards features and judge whether a concerned object/event is detected or not. There are several types of features that can be extracted, such as the state-of-the-art scale invariant feature transform (SIFT) descriptors.

Mobile cameras (e.g., dash cameras or smartphone cameras), thanks to their flexibility, are becoming more and more attractive in object/event detection for smart cities. Due to the mobility nature of mobile cameras, wireless wide area networks (WWAN) have to be used for the video chunk delivery. Therefore, aggressive video compression is required because of the bandwidth-limited and the error-prone nature of wireless networks. To adapt to wireless network capacity, the conventional method is to adjust the quantization parameters (QPs) of videos. However, when the QP is large, the artifact caused by higher compression ratio can significantly distorts the extracted features, which leads to noticeable object/event detection performance degradation (with higher false-positive/negative). As depicted in Fig.~\ref{Fig1}, edge nodes, such as smart phones, can extract features from uncompressed or slightly compressed video chunks. Then, features instead of the full video sequences are uploaded to the IoT servers for the final object/event detection. In this scenario, it is critical to extract and transmit proper features that adapt to wireless resources and surveillance performance requirements.

\subsection{On-Orbit Processing in Satellite Networks}
A satellite network can be consisted of low earth orbit (LEO) satellites, medium earth orbit (MEO) satellites, geosynchronous orbit (GEO) satellites, and ground stations (contain IoT servers). Due to the superior coverage capabilities of satellite networks in communication and imagery, the emerging video satellites can be utilized to perform continuous remote sensing. Video satellites are particularly well suited to track dynamic targets, such as ships and aircrafts. Compared to traditional remote sensing systems, video satellite increases the temporal resolution to a large extent. Thus, there is a trend that video satellites can be widely used in fields like disaster monitoring, resource census, and marine surveillance. Since the capacity of microwave channel between ground station and the satellite is constrained, the spatial resolution of video satellite images is subjected to large compression ratios. As a consequence, super resolution techniques are applied to enhance low resolution video frames to higher resolution ones, in order to satisfy the requirements of some applications~\cite{LJW2017}. Some other studies also proposed pan-sharpening method that yields the high resolution multispectral image (MSI) by merging a low spatial resolution spectral image and a high spatial resolution panchromatic image~\cite{LTX2017}. However, above post-processing methods executed at the ground IoT server cannot perfectly solve the low-resolution problem.

Considering the fact that multiple satellites can perform formation flying and communicate with each other via high capacity optical/microwave links, some of them can share their underutilized processors as edge nodes and provide on-orbit processing capabilities. For the task of target tracking, satellite edge nodes can provide the function of region-of-interest (ROI) slicing. For example, with {\it a priori} knowledge that aircrafts fly along predetermined air routes, edge nodes can slice ROIs that only contain areas around air routes and send ROIs to the IoT servers, which can significantly enhance the Peak Signal to Noise Ratio (PSNR) or Structural Similarity Index (SSIM) of the interested parts of videos under limited space-to-ground bandwidth.

\subsection{Driving Assistance in Internet-of-Vehicles}
Internet-of-Vehicles is one of the revolutions mobilized by IoT, which is evolving from Vehicular Adhoc Networks (VANETs) or vehicular networks~\cite{AAK2017}. Specifically, Internet-of-Vehicles has five types of vehicular communications: V2V, Vehicle-to-Road side unit (V2R), Vehicle-to-Infrastructure of mobile networks (V2I), Vehicle-to-Personal devices (V2P) and Vehicle-to-Sensors (V2S). Driving assistance is an important video-driven task in Internet-of-Vehicles, which leverages cameras, sensors, and edge computing capabilities of On-Board-Unit (OBU) in each vehicle. As an illustrative example, the windshield inbuilt screen of the host vehicle can display preceding vision-blocking vehicles as see-through tabular objects. Such vision-enhancing experience is realized according to captured videos from preceding vehicles, and sensor readings about the surrounding environment. Because of the increased visibility of nearby vehicles, humans and other objects, the process of driving becomes safer.

To reduce the delay of the driving assistance task, OBUs of preceding vehicles can serve as edge nodes and have view transformation functions installed. The view transformation means resizing and adjusting the video captured by the dash camera according to the locations of the host vehicle and the preceding vehicle. Then, instead of original videos, the preceding vehicles send the processed videos to the host vehicle for the low delay driving assistance.

\subsection{General Model of Edge Video Processing}
\begin{figure}[!t]
  \centering
  \includegraphics[width=1.05\linewidth]{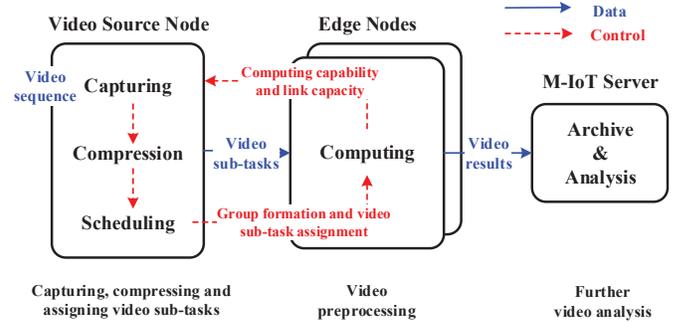}\\
  \caption{General model of edge video processing.}\label{Fig2}
\end{figure}
For edge computing of delay-sensitive video tasks, a video source node (e.g., camera or visual sensor) can offload its video task to nearby edge nodes (e.g., smart phones, satellite processors, or vehicle OBUs) via local wireless/optical networks, and the edge nodes are within the local communication range of the video source node. The video source node captures video sequences (i.e., video tasks), divides each of them into multiple video sub-tasks, compresses these video sub-tasks and delivers them to edge nodes. Next, edge nodes execute video processing functions (e.g., computing SIFT descriptors) with the received video sub-tasks and upload the results to an M-IoT server for further video analysis, such as object/event detection. A delay-sensitive video task is supposed to be processed within a deadline and will fail if the deadline is passed. The general model of the edge video processing is illustrated in Fig.~\ref{Fig2}, which consists of three main components, namely video source node, edge node and M-IoT server~\cite{LCJ2017}.
\begin{itemize}
  \item Video source node: The video source node generates video tasks periodically, divides each video task into a number of video chunks (sub-tasks), compresses video chunks at certain compression ratios and assigns compressed video chunks among all the edge nodes according to scheduling policies.
  \item Edge node: The edge node is with surplus computational ability and storage capacity, helping preprocess video sub-tasks, e.g., feature extraction. Moreover, edge nodes can form cooperative groups based on specific group formation policy and receive the compressed video chunks according to the video sub-task assignment policy.
  \item M-IoT server: The M-IoT server collects the preprocessing results from edge nodes and performs further video analysis, which has abundant computational abilities.
\end{itemize}

Please note that, the video source node can transmit video sub-tasks to edge nodes in multiple transmission modes, for instance, the multicast mode or the unicast mode (see Fig.~\ref{Fig1}). In the multicast mode, a video sub-task is simultaneously transmitted to multiple edge nodes, where these edge nodes can jointly preprocess different parts of this video sub-task for the sake of accelerating the computation. In the unicast mode, a video sub-task is transmitted to only one edge node. Obviously, the realization of edge video processing capabilities is essential for the edge computing-assisted M-IoT systems, although many studies assumed that edge computing functions have already been installed into heterogeneous edge nodes without a real deployment. Recently, authors in \cite{LWB2016,MOR2017} offered methods of embedding edge computing functions into routers and IoT devices via the virtualization technologies. However, they did not focus on video processing tasks for M-IoT systems, and the implementation of cooperative video processing across different edge nodes is not fully investigated. In the next section, we will give an effective solution to the implementation of edge processing functions over an edge computing platform.

\section{Implementation of Edge Processing Functions}

\subsection{Light-Weighted Virtualization}

\begin{figure}[!t]
  \centering
  \includegraphics[width=1.0\linewidth]{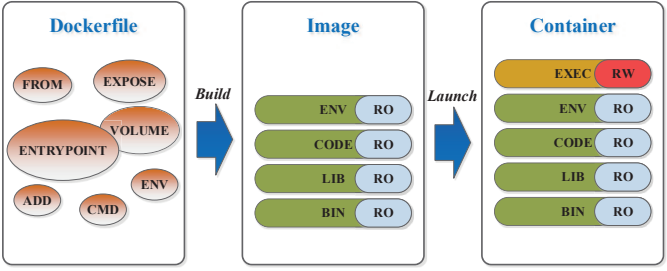}\\
  \caption{An illustration of the Docker virtualization technology.}\label{Fig3}
\end{figure}

For a specific video processing task, edge processing functions may need to be deployed at edge nodes with heterogeneous characteristics and capabilities. One paradigm to ensure that diverse edge nodes perfectly execute the same functions arises from the possibility of using light-weighted virtualization technology, i.e., container technology~\cite{MOR2017}. Containers package code and dependencies together and run on a single host machine based on operating system (OS) virtualization, isolating applications from one another and the underlying infrastructure. Thus, the immigration of applications between different OSs and infrastructures, as well as the configuration of environments, do not cause problems anymore.
Virtual Machine (VM) is another virtualization technique, which is an abstraction of physical hardware. VM includes a full copy of an OS, costing more running space and booting time. Compared to VM that virtualizes hardware, container virtualizes the OS and shares the same OS kernel with the host machine, while the applications in different containers run separately. Therefore, container with lower overhead and faster initiation has a notable advantage over VM in terms of performance in the edge computing.

The proposed edge computing platform leverages the power of Docker, an open source manager of containers~\cite{DOC2017}. To be more concrete, Docker manages the functions, launch, resource allocation and collaborative applications of the containers. As illustrated in Fig.~\ref{Fig3}, a Docker image, built from Dockerfile, serves as a portable container launcher. Docker image features lightweight build, stand-alone operation and facilitated execution, where read-only binaries (BIN), codes (CODE), libraries (LIB), and environment variables (ENV) are included. Any host machine with Docker installed (i.e., a {\it Docker machine}) can be a feasible operating environment for a Docker image. An application is integrated into a layered image built by Dockerfile. Specifically, each Dockerfile command forms a read-only layer of the image. A common layer can be shared among different images. Thus, storage space and download time consumption get minimized. An executing image will launch a container with one read-write execution layer added to the image. In other word, a container is a running instance of an image. A Docker-based application in a container is packaged with corresponding development and staging environments, thus it can run on any Docker machine regardless of conflicts with other applications or establishment of configuration environments. In next subsection, we will further discuss how to realize edge video processing functions in a cooperative manner based on the Docker swarm mechanism.

\subsection{Cooperative Processing Based on Docker Swarm Mechanism}

\begin{figure}[!t]
  \centering
  \includegraphics[width=1.0\linewidth]{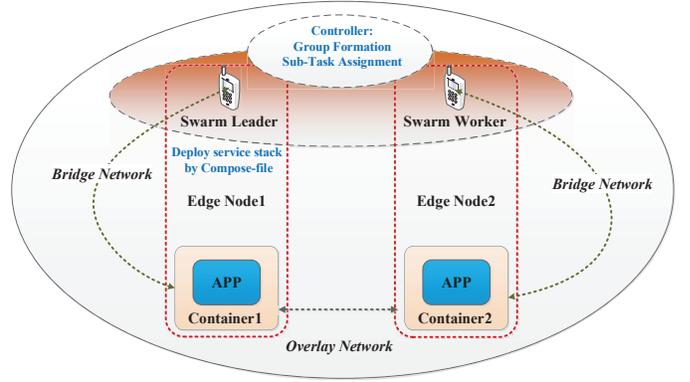}\\
  \caption{An illustration of cooperative processing based on Docker swarm mechanism.}\label{Fig4}
\end{figure}

It is widely observed that edge nodes can be full-loaded or with insufficient computing resources. Thus, offloading the video task to a single edge node may result in unsatisfactory performance for the customers of M-IoT systems. Therefore, the issue arises that how to organize a group of edge nodes, while tasking into account the diverse capabilities of edge nodes, the capacities of different links, and main concern of the edge processing task. For the proposed edge computing platform, we leverage the Docker swarm mechanism, which is able to assemble computing resources of multiple Docker machines. One or more applications (APPs) can be placed in the swarm, where heterogeneous swarm nodes can cooperatively handle offloaded video processing sub-tasks.

As depicted in Fig.~\ref{Fig4}, the proposed edge computing platform can harvest computing resources in multiple edge nodes by establishing containers on them based on the Docker swarm mechanism.
Please notice that, the native Docker does not provide any group formation or workload assignment mechanisms, so we design a {\it controller} for the management of the edge computing, who is in charge of cooperative group (swarm) formation and video sub-task assignment.
Suppose that multiple edge nodes are about to form a swarm according to the group formation policy from the controller, some ports need to be opened first. Specifically, port 2377 (TCP) opens for swarm management, port 7946 (TCP or UDP) opens for node communication and port 4789 (TCP or UDP) opens for the overlay network{\footnote{The overlay network is a user-defined network, which can be adopted for communication between containers running on different edge nodes in a swarm.}}. To set up a swarm, an edge node, as the {\it leader} of the swarm, initiates the swarm with its IP address and generates a string of identifying code. Others edge nodes can join the swarm as {\it workers} by replying the identifying code. When a node launches a container, the connection between the container and its host machine is automatically created by a bridge network. In the bridge network, eth0, the network card of the host machine and docker0, the network card of the container, are connected via a network bridge.

Next, the workflow of setting up an arbitrary edge processing function on the proposed edge computing platform is given. First of all, the leader of the swarm activates a Docker application via a compose-file. A compose-file, usually programmed by a kind of markup language such as YAML, is a configuration file for a Docker application running on different edge nodes in a swarm. In the compose-file, the name of the application/function (e.g., feature extraction, ROI slicing or view transformation), the corresponding image, the starting sequence of the applications, the computing resource allocation, the entry point, the scheduling strategy, the overlay network, and the mounted volume, are all regulated. Notice that, it is unnecessary that every edge node in the swarm has the image to launch a container. In fact, the worker in the swarm can also launch a container as long as the leader in the swarm has the corresponding image. This facilitates the practical implementation of the edge computing to a large extent, as some edge nodes may not have the corresponding image for a video processing task, some of which can consume the storage as much as 2GB or even more. When the leader starts or updates an application by a compose-file, workers in the swarm can be assigned video processing sub-tasks even if the corresponding image does not exist in those nodes. Thanks to the layered structure of images and containers, the communication-resource consumption in the process of container establishing in workers only involves the top layer of the container, i.e., the read-write layer. The total time of video task completion in a swarm can be calculated as
\begin{equation}
t^{Total}=t^{CE}+t^{D}+t^{C}+t^{R}.
\end{equation}
Where, $t^{CE}$ denotes the time for container establishing over all edge node in a swarm, $t^{D}$ denotes the time for video chunk (sub-task) delivery to all edge nodes, $t^{C}$ denotes the time for sub-task handling over all edge nodes, and $t^{R}$ denotes the time for sending back video processing results from all edge nodes to the M-IoT server. How to minimize $t^{Total}$ is a challenging problem, which is out of the scope of this article.

In summary, Docker swarm mechanism can serve as a promising enabler of realizing cooperative processing on edge computing platform for M-IoT systems due to its low overhead for edge nodes, proficient collaboration for task processing, easy initiation and facilitated management.

\section{Performance Evaluation}

\subsection{Experiment Setup}

In this experiment, we arrange two edge nodes in a Docker swarm (cooperative group) running a typical video processing function, namely, feature extraction{\footnote{The video processing function does not limited to the feature extraction. By building a Docker image, any video processing function can be deployed.}} from captured video sequences. Without loss of generality, we assume that the two edge nodes equally share the wireless channel for the video source node to edge node links, and the link capacity between a video source node and an edge node equals the link capacity between two edge nodes. The sample video sequence from a surveillance camera is coded in H.264, with a length of 74s, a resolution of 1280$\times$618, and a frame rate of 30.00fps. The video sequence has a total size of 3.76MB and is divided into two equal video sub-tasks. Both the two edge nodes have i7-4790 3.6GHz CPU and 8GB memory. One of the nodes is the leader and the other one is the worker. The Docker image is built by Python codes and based on the basement layer of Ubuntu 14.04 LTS OS. Some video processing related libraries such as OpenCV, Numpy and Pillow are contained. We use the default feature extraction function in OpenCV, the data analysis function in Numpy, and the image processing function in Pillow. The leader has already stored the Docker image while the worker does not have. The leader and the worker establish containers by the Docker swarm mechanism. The resource usage budget of each container is set as 40\% CPU power of a single edge node and 4GB memory. To continually provide reproducible service, the containers should be able to restart automatically. In the compose-file, we set that the containers are running on all the nodes in the swarm with the restarting interval of several seconds. Since the output data of edge nodes is of much smaller size than that of video chunks, we ignore the time consumption of sending back preprocessing results to the M-IoT server in this experiment.

\subsection{Results and Analysis}

\begin{figure}[!t]
  \centering
  \includegraphics[width=1.0\linewidth]{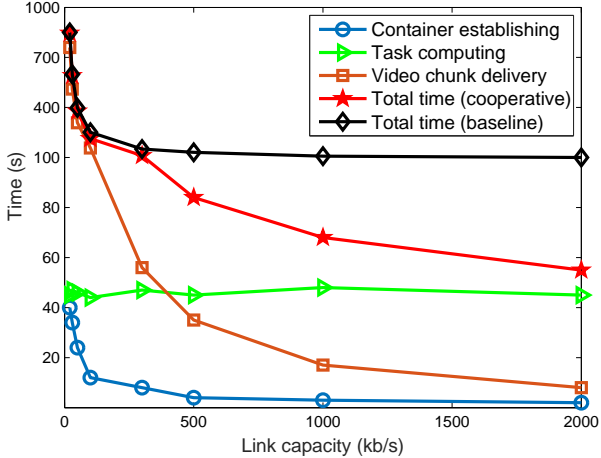}\\
  \caption{Elapsed time under different link capacity from the source video node to one edge node.}\label{Fig5}
\end{figure}

Firstly, we consider a baseline case that the worker is {\it disabled} and only the leader preprocesses the video sequence. That means, both two video sub-tasks are delivered to the leader over the wireless channel and there is no cooperative processing. The total time of baseline is measured and recorded.
Next, for the case of cooperative processing with one leader and one worker (each node preprocesses one sub-task), we measure the elapsed time for container establishing over all nodes, video chunk (sub-task) delivery for all nodes, task computing over all nodes, and the total time of task completion under different link capacity from the video source node to one edge node{\footnote{The link capacity from the video source node to the edge node doubles in the baseline case since only one link occupies the wireless channel.}}, as shown in Fig.~\ref{Fig5}. Notice that the y-axis is not evenly spaced. The step size between 0 and 100 is 20 while the step size between 100 and 1000 is 300. In Fig.~\ref{Fig5}, it is observed that container establishing does consume some time in the cooperative processing case, since the container in the worker is established based on the image in the leader. Generally, Fig.~\ref{Fig5} demonstrates that the container establishing delay and the video chunk delivery delay are influenced by the link capacity, while the computing time consumption is not influenced. Compared to the cooperative processing case with two nodes, the total time consumption of video task completion of the baseline case is much longer. For example, when the link capacity is 1000kb/s, cooperative processing can save 37\% total time compared to the baseline.

In summary, when the link capacity is inadequate (e.g., $<$300kb/s), the time for video chunk delivery is the key determinant of the total time. When the link capacity is adequate, the task computing time is the key determinant of the total time. The cooperative processing is superior to the baseline, especially when the link capacity is sufficient.

\section{Challenges and Opportunities}

Edge video processing for M-IoT systems faces many challenges. The first challenge is that, the management overheads of large scale edge computing-enabled networks could be significant when the number of edge nodes is huge and cooperative nodes are widely spread. The second one is that, the distributed edge computing resources should be carefully scheduled to enlarge the improvement of utilizing edge video processing. The last challenge is that the quality assessment of videos in M-IoT systems is still not well investigated yet.

\subsection{Management of Edge Computing-Enabled Networks}
The current design of edge computing-enabled networks is usually based on the TCP/IP protocol, which is the foundation of Internet. Following the principle of TCP/IP, every node, namely, video source node, edge node or server node, should operate according to IP addresses for edge video processing and cooperative group formation. The management of edge computing-enabled networks involves joint scheduling of communication, computation and storage resources, thus dedicated servers should be deployed as the controller, and the management overheads would be significant if the network is of large scale.

Inspired by the fact that network nodes concern more about what content the requested data contains, rather than where the data come from, Information Centric Networking (ICN) is proposed to primarily achieve efficient information dissemination. In such a network paradigm, the information or content replaces the host as the key network element. Named Data Networking (NDN) is one well-known approach of realizing ICN~\cite{LJW2017}. In general, NDN can be characterized as follows. Firstly, data packets are hierarchically named by their contents, decoupling from the location of the host. Secondly, routers are equipped with caching and computing functions to directly cache and process some contents in the network, without involving the application layer. Thirdly, NDN replaces the traditional channel-based transmission mode with a hop-by-hop one to facilitate efficient multicast. As a result, NDN can natively support distributed resource management through in-network caching and computing capabilities. Thus, the management overheads of edge computing-enabled networks can be cut down compared to centralized management by dedicated controllers. In summary, NDN has the potential to be more suitable for dynamic and large scale scenarios than TCP/IP protocol. It is meaningful to study how to design NDN-like protocols for processing videos of M-IoT systems through widely spread and heterogeneous edge nodes.

\subsection{Intelligent Edge Computing Resource Placement}
Considering the container-based edge computing, resource placement (including Docker images and container resources) and workload assignment have strong impact on the service latency~\cite{WZT2017}, which are essential for delay-sensitive video tasks. The location of edge node that has Docker image stored decides the distance from the source video node to the candidate swarm leader. The variety of images owned by swarm members decides the service scalability of the cooperative group. Besides, the computing and caching resources of each container is finite, so the workload assigned to each container can affect the average response time significantly. It is worth noting that the problems of Docker image placement and container resource management for M-IoT systems, with diverse edge node capabilities, delay-sensitive video tasks, have not been fully considered yet.

\subsection{Quality Assessment of Videos in M-IoT Systems}
For human-viewing videos, Quality-of-Experiences (QoE) is an advanced performance metric for the video quality assessment~\cite{TAO2015}. However, the quality assessment for videos in M-IoT systems is not the same with QoE for human-viewing videos. It is of great importance to design an appropriate video quality assessment scheme for M-IoT systems, which has a main concern of ensuring video processing related performance, such as object detection accuracy and driving safety. It is possible to utilize deep neural network (e.g., Google Tensorflow) as a function approximator of quality score of M-IoT videos due to its universal approximation capability. How to learn correlations between the historical data and predicted video processing quality score is quite critical for the quality assessment of videos in M-IoT systems.

\section{Conclusion}
In this article, we have surveyed the {\it status quo} of edge video processing for M-IoT systems in typical scenarios, such as smart cities, satellite networks, and Internet-of-Vehicles. Then, a general model of edge video processing is summarized. An edge computing platform that supports cooperative video processing has been developed based on the light-weighted virtualization technology, i.e., Docker. The video task completion performance of the proposed edge computing platform has been evaluated, and we observed that cooperative processing outperforms the non-cooperative one. Finally, we provided insights on challenges and opportunities in edge video processing for M-IoT systems.

\bibliographystyle{IEEEtran}
\bibliography{refs}

\end{document}